\documentclass[aps,prd,showpacs,nofootinbib,preprint]{revtex4}
\usepackage{latexsym}
\usepackage{amsmath,amsfonts}
\usepackage{amsbsy}
\usepackage{mathrsfs}
\usepackage{color}

\usepackage{psfrag}

\usepackage{enumerate}

\usepackage{amsmath,amssymb,calc,amsfonts}
\usepackage{latexsym}

\usepackage{graphicx,calc,epsfig}
\def\ut#1{\rlap{\lower1ex\hbox{$\sim$}}#1{}}
\newcommand{\N}{\mathbb{N}}

\newcommand{\be}{\nopagebreak[3]\begin{equation}}
\newcommand{\ee}{\end{equation}}
\newcommand{\ba}{\nopagebreak[3]\begin{eqnarray}}
\newcommand{\ea}{\end{eqnarray}}
\DeclareFontFamily{U}{rsfs}{}         
\DeclareFontShape{U}{rsfs}{m}{n}{<5> rsfs5 <6><7> rsfs7          %
  <8><9><10><10.95><12><14.4><17.28><20.74><24.88> rsfs10}{}     %
\DeclareMathAlphabet{\mathfs}{U}{rsfs}{m}{n}                     %
\newcommand{\mfs}[1]{\mathfs {#1}}                               %
\newcommand{\n}{{\nonumber}}
\newcommand{\va}{\scriptscriptstyle}

\newcommand{\sZ}{{\mfs Z}}

\newcommand{\nn}{\sqrt{j(j+1)}}

\def\pb#1{\rlap{\lower1.5ex\hbox{$\longleftarrow$}}{#1}}
\def\dpb#1{\rlap{\lower1.5ex\hbox{$\Longleftarrow$}}{#1}}
\def\spb#1{\rlap{\lower1.5ex\hbox{$\leftarrow$}}{#1}}
\def\sdpb#1{\rlap{\lower1.5ex\hbox{$\Leftarrow$}}{#1}}

\definecolor{blue}{rgb}{0,0,1}
\definecolor{green}{rgb}{0,1,0}
\definecolor{red}{rgb}{1,0,0}
\definecolor{vio}{rgb}{1,0,1}
\definecolor{ama}{rgb}{1,1,0}

\begin{document}

\title{Statistics, holography, and black hole entropy in loop quantum gravity}

\author{Amit Ghosh}
\affiliation{Saha Institute of Nuclear Physics, 1/AF Bidhan Nagar,
700064 Kolkata, India.}

\author{Karim Noui}
\affiliation{Laboratoire de Math\'ematique et Physique Th\'eorique, 37200 Tours, France;}
\affiliation{F\'ed\'eration Denis Poisson Orl\'eans-Tours, CNRS/UMR 6083;}
\affiliation{Laboratoire APC -- Astroparticule et Cosmologie, Paris 7, 75013 Paris, France.}

\author{Alejandro Perez}
\affiliation{Centre de Physique Th\'eorique, Aix Marseille Universit\'e, CNRS, CPT, UMR 7332, 13288 Marseille, France; 
    Universit\'e de Toulon, CNRS, CPT, UMR 7332, 83957 La Garde, France.}

\begin{abstract}
In loop quantum gravity the quantum states of a black hole horizon are produced by point-like discrete quantum geometry excitations (or {\em punctures}) labelled by spin $j$. The excitations possibly carry other internal degrees of freedom also, and the associated quantum states 
are eigenstates of the area $A$ operator. On the other hand, the appropriately scaled area operator $A/(8\pi\ell)$ is also the physical Hamiltonian associated with the quasilocal stationary observers located at a small distance $\ell$ from the horizon. Thus, the local energy is entirely accounted for by the geometric operator $A$.
We assume that:
\begin{enumerate}
\item  In a suitable vacuum state with regular energy momentum tensor at and close to the horizon the local temperature measured by stationary observers is the Unruh temperature and the degeneracy of `matter' states is exponential with the area $\exp{(\lambda A/\ell_p^2)}$---this is supported by the well established results of QFT in curved spacetimes, which do not determine $\lambda$ but asserts an exponential behaviour. 

\item The geometric excitations of the horizon (punctures) are indistinguishable.

\item In the semiclassical limit the area of the black hole horizon is large in Planck units. 
\end{enumerate}
\noindent It follows that:

\begin{enumerate}
\item Up to quantum corrections, matter degrees of freedom saturate the holographic bound, {\em viz.} $\lambda$ must be equal to $\frac{1}{4}$.

\item Up to quantum corrections, the statistical black hole entropy coincides with Bekenstein-Hawking entropy $S={A}/({4\ell_p^2})$. 


\item The number of horizon punctures goes like $N\propto \sqrt{A/\ell_p^2}$, {\em i.e.}
the number of punctures $N$ remains large in the semiclassical limit.

\item Fluctuations of the horizon area are small ${\Delta A}/{A}\propto \left({\ell_p^2}/{A}\right)^{1/4},$ while fluctuations of the area of an individual puncture are large (large spins dominate). A precise notion of local conformal invariance of the thermal state is recovered in the $A\to\infty$ limit where the near horizon geometry becomes Rindler.

\end{enumerate}
We also show how the present model (constructed from loop quantum gravity) provides a regularization of (and gives a concrete meaning to) the formal Gibbons-Hawking euclidean path-integral treatment of the black hole system. These results offer a new scenario for semiclassical consistency of loop quantum gravity in the context of black hole physics, and suggest a concrete dynamical mechanism for large spin domination leading simultaneously to semiclassicality and continuity.

\end{abstract}


\maketitle

\section{Introduction}

Recent developments in loop quantum gravity (LQG) have brought the problem of black hole physics back to the central  stage. Several pieces of evidence seem to indicate that a clear understanding of these emblematic systems of quantum gravity is perhaps within the reach of LQG. One important input for such new perspective is the quasilocal description of black hole mechanics in terms of stationary observers at proper distance $\ell$ from the black hole horizon \cite{Frodden:2011eb}. Such a treatment leads to an effective notion of horizon energy (suitable for statistical mechanical considerations) in the form
\be\label{energy}
E=\frac{A}{8\pi \ell}.
\ee
It has opened up possibilities of treating the statistical mechanics of quantum {\em isolated horizons} \cite{Ashtekar:2004cn} from the more natural perspectives of {\em canonical} and {\em grand canonical} ensembles \cite{espera}. Another important insight came from the application of {\em spin foam} tools combined with ideas of standard quantum field theory that has provided the means of computing local temperature of quantum horizons \cite{Bianchi:2012ui}.

{ A central input in the present work is the validity of a weak form of holography for `matter' fields (in this paper by matter we mean all non-geometric fields) close to the horizon which is encoded in the assumption that the degeneracy $D(A)$ of matter states for a given horizon area $A$
behaves as
\be\label{holos}
D \propto \exp(\lambda A/\ell_p^2)
\ee
where $\lambda$ is an unspecified dimensionless constant, $\ell_p$ the Planck length. The proportionality constant (or eventually corrections, 
as those considered in Appendix \ref{ap}) 
in (\ref{holos}) is unfixed and for simplicity we assume that $D=\exp(\lambda A/\ell_p^2)$ in the paper. Several calculations in support of such an assumption are available from studies of quantum field theory in curved spacetimes where the statistical properties of matter degrees of freedom in a suitable {\em vacuum} state close to an event horizon are used in the context of local stationary observers. The earliest model is the brick-wall paradigm \cite{'tHooft:1984re, Mukohyama:1998rf}. It is followed by many subsequent analysis attempting to explain the origin of black hole entropy from entanglement of matter states across the horizon \cite{Solodukhin:2011gn}. It is important to point out that, while these treatments are all physically well motivated, they only provide some qualitative answers because the results are infected by UV divergences and other uncertainties (such as number of fields or species) are also present (may be these are also related to the UV-incompleteness of the treatments). The parameter $\lambda$ 
is not fixed by semiclassical QFT considerations (see however, \cite{Bianchi:2012br}). We will use this semiclassical result with an uncertain coefficient $\lambda$ and our analysis will show how these ambiguities disappear when quantum gravity effects are taken into account. 

There are also some recent studies suggesting a relationship between self-dual variables and holography \cite{Frodden:2012nu,Frodden:2012dq}. In these work it is argued by means of an analytic continuation that the dimension of the Chern-Simons Hilbert space---used in modelling quantum isolated horizons \cite{engle, Engle:2010kt}---grows exponentially with $A/(4\ell_p^2)$ for large spin representations. A link between self-duality and thermal behaviour at Hawking temperature is also suggested by \cite{Pranzetti:2013lma}. Even if these claims are quite striking in view of their potential relevance in the context of the present analysis, their validity needs to be tested further as they rely on the existence of a quantum theory that is not completely under control at present, which is quantum gravity with complex variables. 

We will show that holography along with the assumption that geometric excitations of the horizon are indistinguishable particles leads by itself to a remarkable agreement between the `fundamental' loop quantum gravity description and the low energy semiclassical treatments: in view of our analysis, holography becomes a necessary condition for semiclassical physics.

In this paper we explore the statistical mechanical properties of a certain type of non-interactive system that naturally arises in the description of quantum black hole horizons in loop quantum gravity. The basic inputs come from the generic results of quantum gravity and quantum field theories in curved spacetimes. The results show a remarkable consistency of the semiclassical black hole physics with the naive continuum limit of loop quantum gravity. The lessons of this exercise may be far reaching. We hope it could provide new insights into the (elusive) way low energy and semiclassical limit of loop quantum gravity are obtained.}

The paper is organised as follows. Next Section is devoted to the computations of the black hole partition function in both
the canonical and the grand canonical ensembles. We assume the indistiguishability of the punctures introducing first a Gibbs factor before
considering the exact quantum statistics. This allows us to establish an equation of state for the black hole, to recover the $A/4\ell_p^2$
behavior for the black entropy in the semi-classical regime, and to compute other thermodynamical quantities as area/energy fluctuations or
heat capacity. Furthermore, we show that, when combined with the holographic principle, indistiguishability is essential to recover
the expected classical behavior of the black hole. In Section III, we show how the expression of partition functions we found provide
a regularization and gives a concrete meaning to the Gibbons-Hawking Euclidean path integral. We conclude in the last Section by a discussion
and some perspectives.

\section{Implications of indistiguishability}

Perhaps one of the most fundamental lessons of quantum statistics is the indistinguishable nature of particles. This nature is intrinsic to all elementary particles and several non-elementary composites such as atoms, molecules, etc. However, the situation in quantum gravity, especially in the context of LQG, remains uncertain so far. In the context of black holes, it { was argued that their effective fundamental excitations  
are described by punctures located at the horizon}. Although these punctures are quantum mechanical excitations of geometry and their locations are uncertain, they are often treated as distinguishable in their statistical analysis. In this section, we are trying to investigate the possibility of regarding these punctures as indistinguishable quantum excitations of gravity. { The case of distinguishable punctures is briefly discussed in Appendix \ref{BB} where we show how it is inconsistent with semiclassicallity. 


\subsection{Modeling indistiguishability with the Gibbs factor}

There has been some discussions in the past about the issue of statistics of the punctures that define the quantum black hole states \cite{Pithis:2012xw, bhlqg, Krasnov:1996tb}. In this section, we explore the consequences of assuming that punctures are {\em indistinguishable} excitations of the quantum geometry of the horizon. { In our analysis, we also consider as an additional assumption, the validity of a weak form of holography principle, which we will precisely define below}. Before specifying a concrete statistics for these excitations we will model indistinguishability of punctures by introducing the well-known Gibbs factor. In Section \ref{qstat}, we will do a more concrete analysis by committing to fermionic and/or bosonic statistics. { The results remain qualitatively the same for both statistics}. Similar results are expected for anyonic statistics as well.

\subsubsection{Canonical Partition Function}

We are going to study the statistical mechanical properties of quantum IHs. From the framework of LQG, it turns out that the statistical properties of a quantum IH are very well described as a gas of its topological defects, henceforth called {\em punctures}. Using (\ref{energy}), we take the appropriately scaled IH area spectrum \cite{lqg} to be the energy spectrum of the gas
\be\label{area}
\widehat H|j_1,j_2\cdots\rangle=(\frac{\gamma \ell^2_p}{\ell}  \sum_{p} \sqrt{j_p (j_p+1)})\  |j_1,j_2\cdots\rangle
\ee
where $\gamma$ is the Immirzi parameter, $j_p\in \N/2$ is the spin associated with the $p$-th puncture. A quantum IH contains a finite number $N$  of punctures because each puncture carries a minimal area associated with the lowest value for spin $j=1/2$. In this paper we are going to restrict ourselves to a large number of punctures for which a statistical description is well suited.

We are going to treat these punctures as indistinguishable quantum particles. The appropriate way to implement this 
is to make a choice of statistics for these punctures which is what we will do in the following section. {For simplicity, we first
implement their indistinguishability by simply introducing the Gibbs factor $N!$ in the partition function of a system of distinguishable particles obeying Maxwell-Boltzmann statistics}. When compared with the Bose or Fermi statistics, this reproduces the correct behaviour in the high temperature limit. Then, the  canonical partition function with the Gibbs correction factor is given by
\ba\label{zeta}
Q[N,\beta]=\frac{1}{N!}\sum_{\{s_j\}} D[\{s_j\}] \frac{N!}{\prod_j s_j!}\,\prod_j \, e^{-\beta s_jE_j}
\ea
where, following the equation (\ref{area}),  $E_j=\gamma\ell^2_p\nn/\ell$ is the energy of the j-th puncture, $s_j$ is the number of punctures carrying spin value $j$ and $D[\{s_j\}]$ is the number of states associated with the additional (matter or non-geometric \cite{Ghosh:2012wq}) degrees of freedom for a given configuration $\{s_j\}$. The degeneracy factor $D[\{s_j\}]$ is an essential ingredient for successfully implementing the quantum statistics for punctures. { Before arguing why such a degeneracy factor is necessarily present in LQG, we first show its effects in the statistical mechanical properties of the system.} 

For that purpose, we are going to assume (this is the only assumption we will make) that the degeneracy $D[\{s_j\}]$ is maximal in the holographic sense, namely it grows exponentially with $A/\ell_p^2$. As already mentioned in the introduction, both the brick-wall paradigm \cite{'tHooft:1984re, Mukohyama:1998rf} and all other entanglement entropy calculations \cite{Solodukhin:2011gn} imply that qualitatively this must be the case. However, due to regularization ambiguities associated with UV divergencies  one cannot fix the proportionality coefficient in front of $A/\ell_p^2$
 in these approaches. Consequently, we shall leave this coefficient arbitrary and prove a posteriory that it is equal to $1/4$ up to quantum corrections (this is one of the predictions of our analysis). 
 Accordingly, we write the degeneracy as
\be\label{holo}
D[\{s_j\}] = \exp((1-\delta_h)\frac{A}{4\ell_p^2}) = \prod_j \exp{\frac{(1-\delta_h)a_j s_j}{4\ell_p^2}},
\ee
where $\delta_h$ is a free parameter and $a_j=8\pi\gamma\ell_p^2 \sqrt{j(j+1)}$ is the area eigenvalue associated with a single spin $j$ (see Appendix \ref{ap} for the case where there are power law corrections to (\ref{holo})). {As we shall see, (\ref{holo}) implies that} the system is dominated by large spins. Essentially this means that the area spectrum
linearises 
\be
\sqrt{j(j+1)}=j+\frac{1}{2}+{\cal{O}}(1/j)
\ee
and ${\cal O}(1/j)$  terms can be neglected for our purposes. Moreover, none of the results obtained in this paper would depend on the details of the area spectrum. In fact, as we will see later, even in models where the area spectrum is continuous (such as in \cite{Alexandrov:2004fh}), the conclusions of this paper will remain valid. Using (\ref{holo}), equation (\ref{zeta}) becomes
\ba\label{graphy}
Q[N,\beta]=\frac{1}{N!}\sum_{\{s_j\}} N! \prod_j\frac{1}{s_j!}\,e^{-(\beta-\beta_{\va U}+\delta_h \beta_{\va U})s_jE_j}.
\ea 
where $\beta_{\va U}=2\pi\ell/\ell_p^2$ is the inverse Unruh temperature, which represents the local temperature measured by stationary observers at proper distance $\ell$ from the horizon. So the partition function takes the standard old form (except for the additional Gibbs factor) without the degeneracy but with a new effective inverse temperature $\beta-\beta_{\va U}+\delta_h \beta_{\va U}$. 

In \cite{espera} we have studied the statistical mechanical properties of the Boltzmann gas at the Unruh temperature. From these studies we already know that at the semiclassical limit, $\beta$ is close to $\beta_{\va U}$; so we can introduce a new small dimesionless parameter $\delta_\beta$ such that $\beta=\beta_{\va U}(1+\delta_{\beta})$ where $\delta_{\beta}$ must vanish 
in the limit $\ell_p\to 0$ \footnote{Such quantum corrections to Unruh temperature have been suggested in \cite{Pranzetti:2013lma}.}. As a result, the effective inverse temperature becomes
\be\label{betati}
\tilde\beta=(\delta_{\beta}+\delta_h) \beta_{\va U}.
\ee
It will be convenient to use the notation $\delta\equiv \delta_{\beta}+\delta_h$. We will see below that, with the new ingredient of holography and for macroscopic black holes ($A/\ell_p^2\gg 1$), the gas of indistinguishable punctures attains an equilibrium close the Unruh temperature if only if $\delta=\delta_{\beta}+\delta_h\ll 1$, i.e., $\delta$ is a small quantum correction. 

Since $\tilde\beta=\delta\beta_{\va U}$ and $\delta$ becomes a multiplicative factor in the area spectrum in (\ref{graphy}), as we take $\delta\to 0$ we must have $j\to\infty$ so that the combination $\delta j$ remains finite. This { justifies} why the area spectrum becomes effectively linear for large black holes. 

With these inputs the partition function becomes simply
\be\label{key}
Q=\frac{q^N}{N!}
\ee 
with $q$ the partition function for a single puncture
\be
q=\sum\limits^{\infty}_{j=1/2} \exp(-2\pi\gamma\delta j )
=\frac{1}{\exp(\pi\gamma\delta)-1}
\ee
which blows up as $\delta\to 0$, showing that there is no smooth limit { when  $\delta$ approaches $0$}.
This divergence will be crucial in what follows.


\subsubsection{The grand canonical partition function}

In the regime we {consider here}, we do not expect the number of punctures to be strictly conserved. Hence, it is best to use the grand canonical ensemble. This will also become obvious when we introduce suitable statistics for the punctures in the next section.
%



It is clear that in the limit of large $j$, when the LQG area spectrum becomes linear, a single puncture can split into two punctures without changing the area. This means that, in the large area limit, an arbitrary number of punctures can be created or destroyed for a given area. Since the number of punctures cannot be conserved, the chemical potential must vanish. The situation is very much analogous to a system of photons where the photon number is not conserved (in that case a system with a given energy can contain an arbitrary number of soft photons). This statement is strictly true in the regime of linear spectrum and this is precisely the case we should consider when the temperature is close to the Unruh temperature. For that reason we set the fugacity $z=1$ in what follows.

Thus the grand canonical partition function is given by
\ba \sZ[\beta]=\sum_N \frac{q^N}{N!}=\exp(q)\label{zetatis}\ea
The mean energy $U$ { at the inverse temperature} $\beta=\beta_{\va U}(1+\delta_{\beta})$ is
\begin{align}
U&=\frac{A}{8\pi\ell}=-\partial_{\beta}\log(\sZ)\n
\\ & =\pi\gamma T_{\va U}\frac{ \exp{(\pi\gamma\delta)}}{[\exp(\pi\gamma\delta)-1]^2},
\end{align}
where $\delta=\delta_\beta+\delta_h$.
This gives an equation of state relating the area {(equivalently the energy)  with} the parameter $\delta$
\be\label{bi}
\frac{A[\delta]}{4\ell_p^2}=\frac{\pi\gamma\exp(\pi\gamma\delta)}{[\exp(\pi\gamma\delta)-1]^2}.
\ee
The equation of state gives a one-to-one relation between the area and $\delta$, which shows that for large black holes $A/\ell_p^2\gg 1$ we must consider $\delta\ll 1$. 
In fact, for a given large area the equation of state can be solved for $\delta$, giving
\be
\delta = \frac{1}{\pi \gamma} \ln \left( 1 + 2\pi \gamma \frac{\ell_p^2}{A} + 2 \sqrt{\frac{\pi \gamma \ell_p^2}{A}\left(\frac{\pi \gamma \ell_p^2}{A}+1\right)} \right)
\ee
{ which simplifies when $A \gg \ell_p^2$ according to}
\be
\delta=\delta_{\beta}+\delta_h=\sqrt{\frac{4\ell_p^2}{\pi \gamma A}}+{\cal O}\left(\frac{\ell_p^2}{A}\right).\label{deltaarea}
\ee
Although the above equation shows that only the combination $\delta=\delta_h+\delta_\beta$ is small in the large area limit and also in the classical limit $\hbar\to 0$, we can also infer that in these limits $\delta_h\ll 1$ because we already know that $\delta_{\beta}$ is a small quantum correction which also vanishes in these limits. 
In other words, up to corrections that vanish when $\ell_p\to 0$, non geometric degrees of freedom responsible for the degeneracy of the LQG area spectrum saturate the {\em holographic} bound in the sense of (\ref{holo}). This is a non-trivial prediction of LQG because the standard QFT calculations are always infected by ultraviolet divergences and also by the growing number of matter species which together raise doubts about the holographic bound. However, in LQG a detailed balance 
between the matter sector and geometry seems to play an important role so that neither the ultraviolet divergences nor the species problem appear to threaten the holographic principle. Notice that 
we did not impose the parameter $\delta_h$ to be small to begin with---that it turned out to be so is a definite LQG prediction.

Now we can compute the  entropy $S=\beta U+\log\sZ$ which becomes
\be
S_{G}[A]=\frac{A}{4\ell_p^2} \left[1+\delta_{\beta}+2 \sqrt{\frac{\ell_p^2}{\pi \gamma A}}+{\cal O}(\frac{\ell_p^2}{A})\right].
\ee
One can also compute the area/energy fluctuations, $(\Delta U)^2=-\partial_\beta U$, which gives 
\ba
\frac{\Delta U}{U}=\frac{\Delta A}{A}
=\exp(-\frac{\pi\gamma\delta}{2}) [\exp(2\pi\gamma\delta) -1]^{1/2} =
\sqrt{2\pi\gamma \delta}\, + {\cal O}(\delta) \,.
\ea 
Therefore, these fluctuations are small for large black holes, { more precisely $\frac{\Delta A}{A} \simeq \alpha (\ell_p^2/A)^{1/4}$}
where $\alpha$ is a numerical multiplicative factor.
We can also calculate the specific heat. From (\ref{bi})
\be C=-\beta^2\partial_{\beta}U = \frac{2}{\pi\gamma\delta^3 } (1+{\cal O}(\delta))
\ee
which is positive and large. Note that we have always kept $\ell$ fixed, be it while deriving the formula (\ref{energy}) in \cite{Frodden:2011eb} or in all our subsequent calculations. Therefore, in contrast to other treatments the quasi-local formulation of black hole thermodynamics is thermally stable. This can be compared with other stability conditions,
like putting a box around the horizon. We think that a fixed $\ell$ is serving a similar purpose in our case, while the specific heat $C=\Delta E/\Delta T\to\infty$ because $\Delta T\to 0$ for fixed $\ell$.

Another important thing to note is that the physical quantities other than the quasi-local energy (\ref{energy}), such as entropy, energy fluctuations, and specific heat do not depend on the length scale $\ell$ of the local observers. This shows that $\ell$ plays only a fiducial role in our treatment. In the expression of the energy, $\ell$ plays a somewhat similar role to that of a volume $V$ for standard systems---$\ell$ appears in the energy spectrum just like $V$ does. However, in contrast to standard thermodynamic systems, the scale $\ell$ does not appear in the expression of the entropy, specific heat, and area fluctuations---we believe that this is a reflection of the underlying background independence of gravity.  

In the next section we will treat the system of indistinguishable punctures by introducing some suitable statistics. However, the expressions for the entropy and temperature obtained here are not expected to change substantially as far as the leading order terms are concerned. Usually, for non-gravitational systems different statistics give similar results in the limit of small $\beta$ (high temperature regime) because different energy levels become equally likely and the precise forms of distributions become somewhat irrelevant. 
However, in the present case, and as far as the statistics is concerned, the role of inverse temperature is played by $\tilde \beta=\beta-\beta_{\va U}+\delta \beta_{\va U}$. Therefore, high temperature behaviour is achieved in the limit of small $\tilde\beta$ (which close to $\beta_{\va U}$ is equivalent to small $\delta$). Thus, all energy levels become equally likely when the physical temperature  is close to the Unruh temperature. This occurs because of holography: the exponential growth of degeneracy is compensated by the exponential decay of the Boltzmann factor close to the Unruh temperature. Therefore, a naive analogy with non-gravitational systems (as far as the leading order terms in physical quantities are concerned) shows that the gas is as if at an infinite temperature. This is why the use of exact quantum statistics is not expected to affect the leading order terms and, as we shall see, it will not modify the qualitative behaviour of the sub-leading corrections either.

\subsection{Quantum statistics}\label{qstat}

In the previous section we used the Gibbs factor to correct the Boltzmann distribution in order to deal with the gas of indistinguishable punctures. This is an  inexpensive short-cut-method for dealing with the problem. A precise treatment  should instead  use the actual quantum statistics, which is either bosonic or fermionic. From the framework of LQG, as it is not quite obvious what the actual statistics of these punctures would be, we allow for both possibilities and investigate the cases when the punctures are bosonic and fermionic. 
In fact more complex possibilities may arise because the punctures lie on a {\bf two} dimensional surface. We would also like to discuss some of these other possibilities in this section.

\subsubsection{Bosonic/fermionic punctures} \label{wywy}

We assume that all punctures are either bosons or fermions. Another natural choice is that punctures carrying half-integer spins are fermions while punctures carrying integer spins are bosons. But this case can be worked out from the formulas given here. The generic formulas obtained in the case of either purely bosons or purely fermions are also valid for the latter case.

The canonical partition function takes the form
\be
Q_N=\sum_{\{s_j\}}e^{-\sum_j\tilde\beta s_jE_j}
\ee
where $\tilde\beta=\beta-\beta_{\va U}+\delta_h \beta_{\va U}$ and $\sum s_j=N$, as in the Boltzmann case. The grand canonical partition function at $\beta=\beta_{\va U}(1+\delta_{\beta})$ is
\be \sZ=\sum_NQ_N=\prod_j[1\pm\exp(-2\pi\gamma\delta j)]^{\pm 1}\label{thisone}, 
\ee
where $\delta=\delta_h+\delta_\beta$ and the $\pm$ signs describe fermions and bosons respectively. Again, contributions from the linearised spectrum dominates in the limit $\delta\to 0$. Equivalently, for fermions and bosons respectively
\ba \label{logZbosfer}
\log\sZ=\pm\sum_{j=1/2}^{\infty}\log(1\pm e^{-2\pi \gamma\delta  j}).
\ea
The mean energies for fermions and bosons are
\begin{align}
U&=\frac{A}{8\pi \ell}=-\partial_\beta \log\sZ\n\\ & =\frac{\gamma\ell_p^2}{\ell}
\sum_{j=1/2}^{\infty} \frac{j}{\exp(2\pi\gamma\delta j)\pm 1}
=\frac{\gamma \ell_p^2}{2\ell}\sum_{n=1}^\infty \frac{n}{\exp(\pi\gamma \delta n) \pm 1} .
\end{align}
where, in the last sum, the sum runs over integers and not half-integers.
In the limit of small $\delta$ (or equivalently in the limit of large area $A$) the sum can be 
{\it interpreted as a Riemann sum and therefore it can be} approximated by an integral according to:
\be
U = \frac{\gamma \ell_p^2}{2\ell \delta^2} \, \delta \sum_{n=1}^\infty \frac{n\delta}{\exp(\pi\gamma \delta n) \pm 1} \;
\simeq \; \frac{\gamma \ell_p^2}{2\ell \delta^2} \int_0^\infty \frac{x \, dx}{e^{\pi\gamma x} \pm 1} \,.
\ee
Changing the integration variable $x \mapsto x/(\pi\delta)$ in the previous integral, one sees immediately that 
all the relevant parameters can be scaled out and { the area reduces to the following simple expression}:
\be
A=\frac{4\ell_p^2}{\pi\gamma\delta^2}\int_{0}^{\infty}\frac{x\,dx}{e^x\pm 1}
=\frac{\epsilon_\pm\pi\ell_p^2}{3\gamma\delta^2}
\ee
where $\epsilon_\pm=1,2$ for fermions and bosons respectively. Thus, for a fixed large area $A$ we get { the equation of state} $\delta=\ell_p\sqrt{\pi\epsilon_\pm/3\gamma A}$.

{To compute the entropy, one can proceed in a very similar way. One starts by approximating (\ref{logZbosfer}) with an integral} 
 \be
 \log\sZ \simeq \pm\frac{1}{\pi\gamma \delta}\int_{0}^{\infty}\log({1\pm e^{-x}})\,dx
=\frac{\epsilon_\pm\pi}{12\gamma\delta }.
 \ee
 Thus, the entropy  $S=\beta U+\log\sZ$ is given by 
 \begin{align}
 S[A]
 &=\frac{A}{4\ell_p^2}\left[1+\delta_{\beta}+\sqrt{\frac{\pi\epsilon_\pm\ell_p^2}{3\gamma A}}+{\cal O}(\frac{\ell_p^2}{A})\right],
 \end{align}
 where again $\delta_{\beta}$ in the previous formula appears due to the fact that 
 the equation of state only determines the value the combination $\delta=\delta_{\beta}+\delta_h$.

It is also possible that punctures labelled by half-integer spins behave like fermions and punctures labelled by integer spins behave like bosons. A simple adjustment of equations in the previous calculations  would lead to a new value of $\delta$ and entropy in that case. However, as expected the result is qualitatively the same and no special feature occurs. So from the entropy it seems unlikely that one can settle the issue of statistics.

For the case $z=1$, although there is no thermodynamical meaning to the average number of punctures, it is meaningful to calculate the average number of punctures carrying spin-$j$, that is $\langle s_j\rangle$. It has the standard form
\be \label{meansj}
\langle s_j\rangle=[\exp(2\pi\gamma\delta j)\pm 1]^{-1}
\ee
for Fermi and Bose cases respectively. We can also view $\langle s_j\rangle$ as a probability of occurring the spin value $j$ in the sense that 
the larger the value of $\langle s_j\rangle$ is, the more abundant the corresponding spin value $j$ is. Thus we can introduce the probability distribution 
\be\label{prob} p_j=\frac{\langle s_j\rangle}{\sum_{k} \langle s_k\rangle }
\ee
where we omitted the index $\pm$ in the notation $p_j^{\pm}$ for purposes of simplicity.
Using the previous definition one can compute the mean value of the spin contributing to the surface states 
$\langle j \rangle_{\pm}= \sum_j j p_j$ in the two different statistics. The mean value is a priori a complicated function  
but its behavior for small $\delta$ can be expressed easily using the following results:
\ba
\sum_j \langle s_j \rangle & = & \sum_{n=1}^\infty \frac{1}{\exp(\pi\gamma\delta n) \pm 1} 
\approx \frac{1}{\pi\gamma \delta} \int_{\pi\gamma\delta}^\infty \frac{dx}{e^x \pm 1} \\
  & \approx & 
  \left\{
    \begin{split}
    \frac{\log(2)}{\pi\gamma \delta}  \;\;\;\text{(for Bose statistics)}\\ 
    \frac{\log(\delta)}{\pi\gamma\delta} \;\;\;\text{(for Fermi statistics)}
    \end{split}
  \right.
\ea
where here the symbol $\approx$ means that we consider only the leading order term when $\delta$ 
is small. Furthermore, it is straightforward to compute the behavior of $\langle j p_j \rangle$ for small $\delta$ and to show that it is given by
the formulae
\be
\langle j p_j \rangle \approx \frac{\epsilon_\pm}{24(\gamma\delta)^2} 
\ee
As a consequence,  the mean value $\langle j \rangle_\pm$ for Bose and Fermi statistics behaves as follows when $\delta$ is small:
\ba
\langle j \rangle_{-}& \approx  & \frac{\epsilon_- \pi }{24 \gamma  \log(\delta)} \frac{1}{\delta}  \propto \frac{\sqrt{A/\ell_p^2}}{\log(A/\ell_p^2)} \label{unin}\\
\langle j \rangle_{+}& \approx  & \frac{\epsilon_+ \pi}{24 \gamma \log(2)} \frac{1}{\delta}   \propto {\sqrt{A/\ell_p^2}}.\label{dosin} 
\ea
A similar calculation allows to compute the fluctuations 
$\Delta j_\pm = \sqrt{\langle j^2 \rangle_\pm-\langle j \rangle_\pm^2}$ and to obtain their behavior for small $\delta$
which are given by:
\be\label{jb}
\Delta j_-\approx D_+\sqrt{A/\ell_p^2},\quad\Delta j_+\approx D_-\sqrt{A/\ell_p^2}
\ee
where $D_\pm$ are constants.

 Basically, as the area grows the distribution (\ref{prob}) becomes flatter more slowly in the bosonic case than in the case of fermions. That is the why the growth of $\langle j\rangle$ with area in the bosonic case is slower than in the case of fermions. We also see that although the distribution (\ref{prob}) tell us that individually small spins are more likely, the mean value of the spin is large for large area. This is consistent with the viewpoint that the true classical limit is to be taken for $\langle j\rangle\to\infty$, not some individual spin eigenvalue $j\to\infty$. In other words, although the distribution (\ref{prob}) peaks at smaller spin values, larger spin values contribute in a dominant way---this fact is reflected in the large fluctuations $\Delta j_\pm$ around the mean value for large areas. 

\subsubsection{Anyonic statistics}

An interesting question that often appears in the description of quantum isolated horizons is whether the correct statistics for punctures should be anyonic. This comes from the fact that the horizon degrees of freedom in $2+1$ dimensions are described by a Chern-Simons theory in which the braiding of sources occurs\footnote{See \cite{Sahlmann:2011xu}
for an argument leading to a non trivial statistic of punctures within the LQG framework.}. As a result a new form of statistics appears, called anyonic statistics. Such non-trivial statistics can be heuristically thought of as  a result of a (non-local) quantum interaction. Thus, it is hard to make analytic computations with anyons. Nevertheless,  standard cluster expansion techniques (for standard short range interactions these are excellent approximations in the high temperature and low density regime) have been applied to compute up to the second virial coefficients. These computations show explicitly that such coefficients interpolate between those for Bosons and Fermions depending on the value of a suitable deformation parameter.
For black hole models such deformation parameter is the level of the Chern-Simons theory which grows with the horizon area in Planck units. For macroscopic black holes  one is really in the large level regime where anyons should be well approximated by standard statistics. These points suggest that the qualitative structure of the results presented here should be preserved even in the case of anyonic statistics.

Notice, however, that it is not clear if anyons are physically relevant for black hole physics in the context of loop quantum gravity. The reason is that in LQG the effective $2+1$ dimensional quantum description of the horizon should really be rooted in a $3+1$ description. Therefore, it is by no means clear why {\em braiding} of punctures should play a fundamental role when in LQG when the notion of braiding is trivial from a four-dimensional perspective. 
   
 \section{Semiclassical consistency: general relativity emerging from LQG} \label{sc}


Our expression for the grand canonical partition function in quantum statistics
\begin{align}
\sZ[\beta]&=\sum_N\sum_{\{s_j\}}e^{-\sum_j(\beta-\beta_{\va U}+\delta \beta_{\va U}) s_jE_j}
\n\\&
\approx \sum_N\sum_{\{s_j\}}e^{-(\beta-\beta_{\va U})\sum_js_j\frac{a_j}{8\pi\ell_p^2\ell}}
,\label{gpf}
\end{align}
where in the last line we have neglected the quantum correction $\delta$ in (\ref{betati}). In such regime of small $\delta$ (or equivalently to large horizon area) we may expect the partition function (\ref{gpf}) to arise from some semiclassical considerations also. In the following we will see that this is exactly what happens. 

To explicitly see this, we first view the expression (\ref{gpf}) as a regularized version of some euclidean functional integral where the partition function is obtained by integrating over euclidean metrics $g^{\va(4)}$ with some appropriate weights that depend on the metric $g^{\va(4)}$. Since in (\ref{gpf}) we are summing over all punctures and spin configurations associated with a two-surface (the horizon), it is natural to interpret the sum as a discretization of the continuum integral over the euclidean metrics of the two-surface $g^{\va(2)}$ with the weight factor $\sum s_ja_j=A$ as area $A[g^{\va(2)}]$ of the two-surface associated with the metric $g^{\va(2)}$ 
\begin{align}\label{zetati}
\sZ[\beta]&=\int Dg^{\va(2)}\exp\Big[-(\beta-2\pi\ell)\,\frac{A[g^{\va(2)}]}{8\pi\ell_p^2\ell}\,\Big]\n\\
&\approx\exp\Big[-(\beta-2\pi\ell)\,\frac{A[g^{\va(2)}]}{8\pi\ell_p^2\ell}\,\Big]
\end{align}
In fact, (\ref{gpf}) provides a LQG definition for (\ref{zetati}). Now it is well-known that the expression (\ref{zetati}) can be obtained from a functional integral over a euclidean four-metric $g^{\va(4)}$ in the leading semi-classical approximation (for a fixed background metric $g^{\va(4)}_0$)---this was explicitly shown in \cite{Gibbons:1976ue} for the case where one gives boundary conditions at infinity. In all generality the expression of the partition function is
\begin{align}\label{GH}
\sZ[\beta]
&=\int Dg^{\va (4)}_{\beta} \exp\Big[-\frac{1}{16\pi\ell_p^2}\int\limits_MR[g^{\va(4)}]\n\\ &-\frac{1}{8\pi \ell_p^2}\int\limits_{\partial M}(K-K_0)\,\Big],
\end{align}
where  the index $\beta$ in $g_{\beta}^{\va(4)}$ reminds us that in the semiclassical Euclidean prescription one considers metrics with a conical deficit angle $2\pi-\beta/\ell$ at the horizon, $R$ denotes the Ricci scalar, $K$ is the extrinsic curvature at the boundary $\partial M$ and $K_0$ is a standard counter-term that is to be subtracted for consistency. 

The expression (\ref{zetati}) follows from (\ref{GH}) when one evaluates the functional integral to a spacetime region contained between a stationary black hole horizon and the world-sheet of stationary observers at proper distance $\ell$ used in \cite{Frodden:2011eb}. Then $K$ is the extrinsic curvature of the boundary $\partial M$ and the counter term $K_0=\ell^{-1}$ so that the result is trivial in the $A\to \infty$ limit where the spacetime region is isomorphic to (Euclidean) Rindler spacetime (see Figure \ref{M4}; in fact $K_0$ is the value of the extrinsic curvature on the boundary of that region in the euclidean continuation).  Under such conditions the boundary integral  in (\ref{GH}) falls off like $\ell^3/(\sqrt{A})$ and can be neglected. Therefore, (\ref{zetati}) entirely comes from the curvature term and the conical singularity at the horizon. For more details of the derivation see \cite{thesisE}.

\begin{figure}[h]
 \centerline{\hspace{0.5cm} \(
\begin{array}{c}
\psfrag{a}{$\ell$}
\psfrag{b}{$\ell$}
\includegraphics[height=4cm]{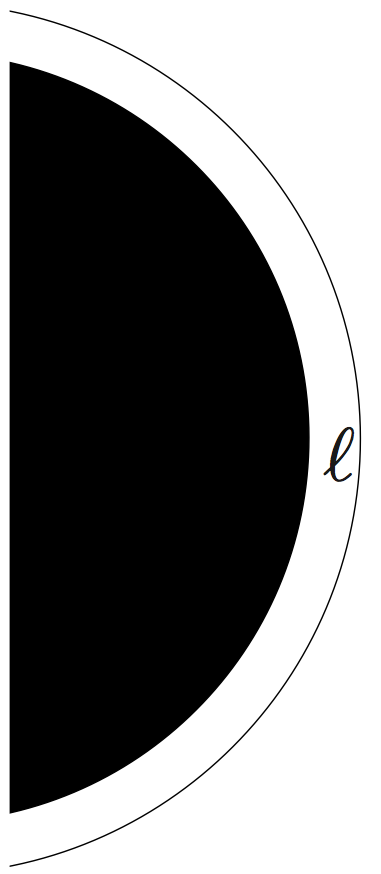}
\end{array} \ \ \ \ \ \ \ \ \ \ \ \ {\LARGE {\to_{\!\!\!\!\!\!\!\!\!\!\!\!{}_{{}_{A\to \infty}}}}} \ \ \ \ \ \ \ \ \ \ \ \ \begin{array}{c}
\psfrag{a}{$\ell$}
\psfrag{b}{$\ell$}
\includegraphics[height=4cm]{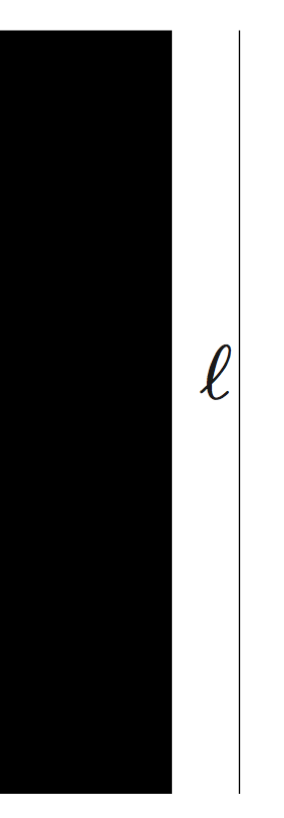}
\end{array}
\)}
\caption{In the infinite area limit one recovers flat spacetime in the local formulation.  The path integral is defined in the portion of spacetime between the horizon and the worlsheet of stationary observers at distance $\ell$.} \label{M4}
\end{figure}

Note that (\ref{GH}) does not depend on the detailed definition of the path integral. It rather follows from the semi-classical approximation. Since (as argued before) in our case large spins dominate (analogue of the semiclassical approximation), we expect that the semi-classical approximation of the spin foam transition amplitudes \cite{han, Barrett:2010ex} would provide a rigorous implementation of this last formal step. This is related to the ideas put forward in \cite{
Bianchi:2012ui} (see version 1 on Arxiv) when exploring the dynamics of a single plaquette (puncture). 

Nevertheless, this analysis shows that the holographic behaviour (\ref{holo}) along with the large area approximation is necessary for the emergence of general relativity from LQG in the context of black holes. A similar point was also discussed  in \cite{Frodden:2012dq}. It provides a semiclassical consistency of LQG at least in this restricted context. Also note that since the statistical physics of a black hole (as described by a gas of punctures) is (in the $A/\ell_p^2\to \infty$ limit) the same as statistical physics in flat spacetime as seen by Rindler observers, the arguments presented here apply to more general cases than just black holes and thus provides new hints as to how to recover standard QFTs in flat space physics in the LQG formalism. Our study shows the relevance of (\ref{holo}) in LQG and its close connection to the semiclassical consistency of LQG.

\section{Discussion}

Let us conclude with the discussion of various aspects of the analysis presented in this paper.

\subsubsection*{Indistinguishability or not indistinguishability?} 

We have shown that the naive introduction of holographic degeneracy of matter states associated with a quantum horizon leads to a whole body of interesting phenomenology. This strongly suggests this degeneracy is a necessary ingredient for the continuum and semiclassical limit of LQG. There are several independent quantum field theoretic scenarios from which such holographic degeneracies (\ref{holo}) can be inferred. However, various issues attached to this holography indicate  that a complete and satisfactory analysis of degeneracy can only be made in a full quantum theory of gravity. One famous example is the brick-wall model of 't Hooft \cite{'tHooft:1984re}. 
The computations of black hole entropy based on matter enganglement are closely related to this example and produce results that are qualitatively same \cite{Solodukhin:2011gn}. One limitation of these  analysis is that the computations are semiclassical (QFT on a classical background spacetime) and thus the regularization ambiguities preclude the precise calculation of the proportionality constant in front of $A/\ell_p^2$ in (\ref{holo}). Moreover, the answer are typically proportional to the number of matter fields considered; this is referred to as the {\em species problem}. 
The holographic degeneracy (\ref{holo}) in all these cases is associated with matter or non-geometric degrees of freedom close to the horizon. Another example (which from a physical viewpoint remains unclear at the moment) is the holographic degeneracy \cite{Frodden:2012dq} found by analytic continuation from real to complex Ashtekar variables. The validity of such an analytic continuation depends on the possibility of defining the quantum theory in terms of complex self-dual variables which at this stage remains an open issue. 
 
Our analysis shows that thermal equilibrium at Hawking temperature implies that as soon as quantum geometry effects are brought in (like, the area spectrum), degeneracy of the area spectrum due to non-geometric degrees of freedom (that for simplicity are here referred to as matter and they include any other excitation that does not affect the horizon area eigenvalues) grows exponentially saturating the holographic bound up to quantum gravity corrections. Our treatment removes the two central problems mentioned above---one, the UV divergences of QFTs (because of the underlying discreteness of LQG) and two, the ambiguities of the proportionality factor, also known as the species problem. However, these remarkable results are obtained here in an indirect way though the inputs (\ref{holos}). We hope that a detailed analysis of the non-geometric degrees of freedom in the framework of LQG would reproduce this result. We expect this to arise from the dynamical restrictions on non-geometric degrees of freedom imposed by the Hamiltonian constraint in combination with the requirement that the suitable quantum gravity physical state describing the BH system approximates the near horizon geometry of a stationary black hole in the semiclassical regime.

 
In this work we have strongly used indistinguishability of LQG excitations of the horizon geometry, i.e., {\em punctures}.
We strongly believe that this is the correct way to treat punctures at the horizon. We base our assumption on one of deepest implications of quantum mechanics about the nature of particles. In a relativistic quantum field theory, particles are excitations of an underlying field (like waves on the surface of a lake). Such excitations by their very nature are indistinguishable. Then consistency (in the standard quantum field theoretic context) requires that these excitations are either fermionic of bosonic (or more exotic possibility like anyonic in lower dimensions). But indistinguishability is a more fundamental issue. From the analysis of the present paper we see that semiclassical consistency in the context of black hole physics implies the necessity of indistinguishability in a strong way (see also Appendix \ref{BB}). 

This seems to contradict some results of the past where punctures were taken to be distinguishable \cite{Ashtekar:2000eq}. We believe that even though the statements made in \cite{Ashtekar:2000eq} are correct in their context, their validity is only a peculiarity of the effective models where such results were derived. Moreover, in such models where no holographic feature (\ref{holo}) is present among other things because matter degrees of freedom are completely ignored, BH entropy will not grow linearly with $A/\ell_p^2$ if punctures are taken to be indistinguishable \cite{Pithis:2012xw, Krasnov:1996tb}. It is only when one includes the effect of matter through (\ref{holo}) that indistinguishability leads to the correct leading-order entropy and the correct leading order temperature. 

\subsubsection*{Matter vs. Geometry} 
 
For quite some time there has been a tension in the field as to what could be the correct source for the huge number of degrees of freedom that lead to black hole entropy for non-extremal black holes. Some argue that it is of a purely geometric origin, while others argue that it entirely comes from matter entanglement close to the horizon. Examples of the first are either the heuristic derivations based on the Euclidean path integral approaches or the more precise but older entropy calculations in LQG that completely neglect matter fields. Examples of the second type are the brick wall models and entanglement entropy calculations. In the scenario presented here, both sources contribute. The geometric component coming from the quantization of the area in LQG and the non-geometric component that produces the degeneracy (\ref{holo}) of the area spectrum. In this paper the two aspects are seamlessly combined which produces results that are compatible with the semiclassical regime. 
 
 \subsubsection*{Continuum and semiclassical limit of LQG} 

From a broader perspective the results presented here might also clarify the difficult issue of how one is expected to recover the continuum and/or semiclassical limit of LQG from a fundamentally combinatorial structure of the theory. In this respect also there have been different views. On one hand, there is an idea that the continuum limit of LQG is to be obtained through physical states constructed out of the very fine grained structures, i.e., spin network graphs with many edges and nodes per unit volume. According to this view the geometric quantum numbers (the spin labels) colouring spin networks seem to be small. On the other hand, investigations in the context of spin foams \cite{Perez:2012wv} indicate  that the basic building blocks of quantum geometry admit a classical geometric interpretation when coloured by large spins. Moreover, it is in the large spin limit where spin foam amplitudes can be related to general relativity (via a Regge regularization) in the semiclassical regime \cite{han, Barrett:2010ex}. 
 {From the present analysis of Section \ref{wywy}---equations (\ref{unin}) and (\ref{dosin})---we find \be
  \langle N\rangle_-\propto \sqrt{A/\ell_p^2}\ {\log(A/\ell_p^2)},\ \ \ \ 
  \langle N\rangle_+\propto \sqrt{A/\ell_p^2}\ee
  for Bose and Fermi statistics respectively.
 These numbers as well as (\ref{unin}) and (\ref{dosin}) are all large in the macroscopic regime and thus, the tension evoked above disappears.}
  Thus, the system is dynamically driven to the configurations one would expect 
 from naive semiclassical and continuum limit considerations. Notice that the thermal state of the
 system respects the hierarchy of scales
 \be
 \ell_p^2\ll \bar a=\ell_p^2 \langle j \rangle\ll A
 \ee 
 that is found to correspond to the semi-classical and low energy limit in 
 \cite{Sahlmann:2001nv}
 and coincides with the analog criterion found in the context of spin foams \cite{han}.
 Perhaps not surprisingly this is complemented by the clear-cut 
 link with the semiclassical Euclidean  semiclassical formulation as described in Section \ref{sc}.
 
 \subsubsection*{IR vs. UV Newtonian constants}

For simplicity of presentations here we have not explicitly considered the possibility that the value of Newton's constant $G$ appearing in the fundamental (or UV) area spectrum in units of $\ell_p$ could be different from its low energy or large scale value. As in any quantum field theory, $G$ is expected to flow with a suitably defined phenomenological scale. Even though the explicit way such a flow would occur cannot be addressed in the framework of the present calculation; one can explore the possible modification of our results in the case where we assume that $G_{\va UV}\neq G$, where $G_{\va UV}$ and $G$ are the fundamental high energy and low energy values of the gravitational constants respectively. This would translate into two different values of the Planck lengths $\ell_p^{\va UV}\neq\ell_p$. In particular, $\ell_p^{\va UV}=\sqrt{\hbar G_{\va UV}}$ will appear in the fundamental area spectrum while $G$ will appear in the form of the effective Hamiltonian. Explicitly, equation (\ref{area}) becomes
\be\label{areap}
\widehat H|j_1,j_2\cdots\rangle=(\frac{ \hbar \gamma }{ \ell}\frac{G_{\va UV}}{G}  \sum_{p} \sqrt{j_p (j_p+1)})\  |j_1,j_2\cdots\rangle
\ee 
Notice that the change with respect to (\ref{area}) can be hidden by the shifting the Immirzi parameter  $\gamma \to \gamma G/G_{_{\va UV}}$. By just looking at the dependence of our results on the Immirzi parameter one concludes that the flow of Newtons constant does not affect the leading order result of our calculations. In particular, the low energy Newton's constant $G$ appears in the leading term of the black hole entropy and one genuinely recovers the approapriate form of Bekentein-Hawking entropy formula suggested by the classical first law.

\subsubsection*{Conformal invariance at Unruh temperature}

Notice that in the limiting case $A/\ell_p^2\to\infty$, $\delta\to 0$ and the spacetime close to the horizon becomes flat (this is the Rindler limit, the limit of zero curvature). The probability distribution (\ref{prob}) becomes independent of the value of $j$. This is like a {\em local} scale invariance, local because the above statement holds at each puncture independently and scale independent because the spacetime close to the horizon possesses no intrinsic scale. 
Is this the origin of an underlying conformal invariance advocated in some approaches \cite{Carlip:2011vr, Dreyer:2013noa}? At first sight this seems to be different because the conformal invariance that emerges in such treatments are associated with the $r-t$ plane while here the scale invariance appearing in our context is on the two surface of the horizon. Nevertheless, the mechanism that makes the results of these treatments generic might be also relevant here for similar geometric reasons. This is an important issue that is to be clarified by further analysis.

\section{Acknowledgments} 

We are grateful to  E. Frodden, M. Geiller, M. Han, and D. Pranzetti for exchanges and discussions contributing to this work.
A.P. thanks the quantum gravity group at Marseille for inputs and discussions during an informal
presentation this work; specially to C. Rovelli for discussions and support.

\begin{appendix}
\section{Logarithmic corrections}\label{ap}

In (\ref{holo}) we have assumed only the leading asymptotic growth of the density of matter states. The leading term is often accompanied by a power law suppression leading to an expression for the degeneracy of the form 
\begin{align}\label{lolo}
D[\{s_j\}]&\approx(A/\ell_p^2)^{\alpha}\exp((1-\delta)A/(4\ell_p^2))\n\\ &=(\sum_{j} a_j s_j)^{\alpha}\prod_j\exp{\frac{(1-\delta)a_j s_j}{4\ell_p^2}},
\end{align}
where $\alpha$ is in general an arbitrary real number (usually negative). For simplicity let us choose $\alpha=k$ where $k$ is a positive integer. In such case the partition function becomes higher derivative of (\ref{zetatis})
 \ba \sZ[\beta]=\omega(-\partial_{\tilde\beta})^k\exp(q)=\omega(T_{\va U})^k(-\partial_\delta)^k\exp(q)\label{zetate}\ea
where in the following discussion the exact form of the pre-factor $\omega$ will not be needed. From (\ref{zetatis}) and considering small $\delta$ or equivalently large $q$, $(-\partial_\delta)e^q\approx(\pi\gamma)q^2e^q$. Therefore, to the leading order 
\begin{align}
\log\sZ[\delta]&=q+2k\log q+o(1/q)\n\\ &\approx\frac{1}{\pi\gamma\delta}-2k\log(\delta)+o(\delta).
\end{align}
From the above formula, it is easy to see that the mean energy is not affected to the orders we are considering now
\begin{align}
U&=\frac{A}{8\pi\ell}=-\partial_{\beta}\log\sZ\n\\ 
&=\frac{\ell_p^2}{2\pi^2\ell\gamma\delta^2}[1+o(\delta)].
\end{align}
Therefore, the relation (\ref{deltaarea}) between $\delta$ and area is not modified to leading orders. However, the entropy $S=\beta U+\log\sZ$ receives a logarithmic correction
\ba\label{logcorrect}
S_{G}[A]&=&\frac{A}{4\ell_p^2} \Big[1+\delta_{\beta}+4 \sqrt{\frac{\ell_p^2}{\pi \gamma A}}\Big]\n \\  &+& k\log\left(\frac{A}{\ell_p^2}\right)+o(1).
\ea
Thus, the logarithmic corrections to the entropy are associated to the power-law suppression of the exponential growth of states (\ref{holo}). The entropy formula (\ref{logcorrect}) can be extended to some negative real number $\alpha$ by analytic extensions. This shows that the mechanism leading to the logarithmic corrections by loop corrections investigated in \cite{Sen:2012dw}
will also lead to the very same contributions in our analysis. This is due to the fact that such logarithmic corrections would modify (\ref{holo}) according to (\ref{lolo}).

\section{Distinguishable punctures}\label{BB}

It is easy to see that the distinguishable nature of the punctures is not
consistent with holography (in the precise sense we have implemented
holography in this paper). Close to the Unruh temperature the grand
canonical partition function is given by $\sZ_{\rm MB}=\sum_{N}q^N$. Its 
convergence requires $q<1$, which means $\delta > \log 2/(\pi \gamma)$, 
hence a deviation from holographic bound is expected.

One could try to ignore this and analytically continue the series from its
value for $|q|<1$ and write
\ba
\sZ_{\rm
MB}=\sum_{N}q^N:=\frac{\exp(\pi\gamma\delta)-1}{\exp(\pi\gamma\delta)-2}.
\ea
However, the above partition function becomes negative for small $\delta$
and cannot be used to construct a real thermodynamical potential
$\log\sZ_{\rm MB}$.

Distinguishability of punctures seems to be incompatible also with holography in the sense of having a factor close to $1/4$ in (\ref{holo}). But it would be compatible with a prefactor close to $(\pi\gamma-\log(2))/(4\pi\gamma)$ in this case. However, a simple calculation shows that if that were the case, then the entropy is not close to Hawking entropy and the correspondence between the statistical mechanical partition function and the Euclidean path integral semiclassical 
partition function would be lost.

\end{appendix}


\begin{thebibliography}{1}

\bibitem{Hawking:1974sw}
  S.~W.~Hawking,
  Commun.\ Math.\ Phys.\  {\bf 43} (1975) 199.
 \bibitem{Frodden:2011eb}
  E.~Frodden, A.~Ghosh and A.~Perez,
  arXiv:1110.4055 [gr-qc].

\bibitem{Bianchi:2012ui}
  E.~Bianchi,
  arXiv:1204.5122 [gr-qc].
  E.~Bianchi and W.~Wieland,
  arXiv:1205.5325 [gr-qc].

\bibitem{'tHooft:1984re}
  G.~'t Hooft,
  Nucl.\ Phys.\ B {\bf 256} (1985) 727.
  
  \bibitem{Mukohyama:1998rf}
  S.~Mukohyama and W.~Israel,
  Phys.\ Rev.\ D {\bf 58} (1998) 104005,
  arXiv:gr-qc/9806012.

  \bibitem{Solodukhin:2011gn}
  S.~N.~Solodukhin,
  Living Rev.\ Rel.\  {\bf 14} (2011) 8,
  arXiv:1104.3712 [hep-th].

\bibitem{Bianchi:2012br}
  E.~Bianchi,
  arXiv:1211.0522 [gr-qc].


\bibitem{Frodden:2012nu}
  E.~Frodden, M.~Geiller, K.~Noui and A.~Perez,
  JHEP {\bf 05} (2013) 139,
  arXiv:1212.4473 [gr-qc].

\bibitem{Frodden:2012dq}
  E.~Frodden, M.~Geiller, K.~Noui and A.~Perez,
  arXiv:1212.4060 [gr-qc].


\bibitem{engle} J.~Engle, A.~Perez and K.~Noui,
  Phys.\ Rev.\ Lett.\  {\bf 105} (2010) 031302,
  arXiv:0905.3168 [gr-qc] 
  
  \bibitem{Engle:2010kt}
  J.~Engle, K.~Noui, A.~Perez and D.~Pranzetti,
  Phys.\ Rev.\ D {\bf 82} (2010) 044050,
  arXiv:1006.0634 [gr-qc].

 
 \bibitem{Pranzetti:2013lma}
  D.~Pranzetti,
  arXiv:1305.6714 [gr-qc].
  
\bibitem{bhlqg}
 C.~Rovelli,
  Phys.\ Rev.\ Lett.\  {\bf 77} (1996) 3288,
  arXiv:gr-qc/9603063.
  A.~Ashtekar, J.~Baez, A.~Corichi and K.~Krasnov,
  Phys.\ Rev.\ Lett.\  {\bf 80} (1998) 904,
  arXiv:gr-qc/9710007.


\bibitem{Krasnov:1996tb}
  K.~V.~Krasnov,
  Phys.\ Rev.\ D {\bf 55} (1997) 3505,
  arXiv:gr-qc/9603025.


\bibitem{Corichi:2009wn}
    A.~Corichi,
  arXiv:0901.1302 [gr-qc].
  R. K. Kaul and P. Majumdar, Phys.\ Lett.\ {\bf B439} (1998) 267, arXiv:gr-qc/9801080.
M. Domagala and J. Lewandowski, Class.\ Quant.\ Grav. {\bf 21} (2004) 5233, arXiv:gr-qc/0407051.
K. A. Meissner, Class.\ Quant.\ Grav.\ {\bf 21} (2004) 5245, arXiv:gr-qc/0407052.
A. Ghosh and P. Mitra, Phys.\ Lett.\ {\bf B616} (2005) 114, arXiv:gr-qc/0411035.
   R.~Basu, R.~K.~Kaul and P.~Majumdar,
  Phys.\ Rev.\  D {\bf 82} (2010) 024007, 
  arXiv:0907.0846 [gr-qc].
J. Engle, K. Noui, A. Perez and D. Pranzetti,
JHEP {\bf 1105} (2011) 016, arXiv:1103.2723 [gr-qc].
     

%


\bibitem{Ashtekar:2004cn}
  A.~Ashtekar and B.~Krishnan,
  Liv. Rev.\ Rel.\  {\bf 7}, 10 (2004), arXiv:gr-qc/0407042.

  
 \bibitem{espera} 
  A.~Ghosh and A.~Perez,
  Phys.\ Rev.\ Lett.\  {\bf 107}, 241301 (2011),
  arXiv:1107.1320 [gr-qc].


\bibitem{Krasnov:1997yt}
  K.~V.~Krasnov,
  Class.\ Quant.\ Grav.\  {\bf 16} (1999) 563,
  arXiv:gr-qc/9710006.

\bibitem{countings}
  A.~Ghosh and P.~Mitra,
  Phys.\ Rev.\  D {\bf 74} (2006) 064026.
  arXiv:hep-th/0605125.
   I.~Agullo, J.~F.~Barbero G., J.~Diaz-Polo, E.~Borja and E.~J.~S.~Villasenor,
  Phys.\ Rev.\ Lett.\  {\bf 100} (2008) 211301,
  arXiv:0802.4077 [gr-qc].



\bibitem{G.:2011zr}
A. Chatterjee and P. Majumdar, Phys.\ Rev.\ Lett.\ {\bf 92} (2004) 141301.
  J.~F.~B.~G. and E.~J.~S.~Villasenor,
  arXiv:1106.3179 [gr-qc].

\bibitem{lqg}
T. Thiemann, ``Modern Canonical Quantum General Relativity''
Cambridge, UK: CUP. C. Rovelli,
    `` Quantum gravity,'' Cambridge, UK: CUP.

\bibitem{Alexandrov:2004fh}
  S.~Alexandrov,
  gr-qc/0408033.


\bibitem{Ashtekar:2000eq}
  A.~Ashtekar, J.~C.~Baez and K.~Krasnov,
  Adv.\ Theor.\ Math.\ Phys.\  {\bf 4} (2000) 1,
  [gr-qc/0005126].

\bibitem{Pithis:2012xw}
  A.~G.~A.~Pithis,
  Phys.\ Rev.\ D {\bf 87} (2013) 084061,
  arXiv:1209.2016 [gr-qc].

  \bibitem{Carlip:2011vr}
   S.~Carlip,
  AIP Conf.\ Proc.\  {\bf 1483} (2012) 54,
  arXiv:1207.1488 [gr-qc].
  S.~Carlip,
  Entropy {\bf 13} (2011) 1355,
  arXiv:1107.2678 [gr-qc].
  S.~Carlip,
  JHEP {\bf 1104} (2011) 076,
   [Erratum-ibid.\  {\bf 1201} (2012) 008],
  arXiv:1101.5136 [gr-qc].

\bibitem{Dreyer:2013noa}
  O.~Dreyer and A.~Ghosh,
  arXiv:1306.5063 [gr-qc].
\bibitem{Sahlmann:2011xu}
  H.~Sahlmann,
  Phys.\ Rev.\ D {\bf 84} (2011) 044049,
  arXiv:1104.4691 [gr-qc].

  \bibitem{Gibbons:1976ue}
  G.~W.~Gibbons and S.~W.~Hawking,
  Phys.\ Rev.\ D {\bf 15} (1977) 2752. 
  D.~V.~Fursaev and S.~.N.~Solodukhin, Phys.\ Rev.\ D {\bf 52} (1995) 2133.

\bibitem{thesisE}
  E. Frodden,
  Ph.D. Thesis

\bibitem{Ghosh:2012wq}
  A.~Ghosh and A.~Perez,
  arXiv:1210.2252 [gr-qc].


\bibitem{Frodden:2012dq}
  E.~Frodden, M.~Geiller, K.~Noui and A.~Perez,
  arXiv:1212.4060 [gr-qc].

\bibitem{han}
M.~Han,
  arXiv:1304.5628 [gr-qc].
 M.~Han,
  arXiv:1304.5627 [gr-qc].
 M.~Han and M.~Zhang,
  arXiv:1109.0499 [gr-qc].
 M.~Han and T.~Krajewski,
  arXiv:1304.5626 [gr-qc].

\bibitem{Barrett:2010ex}
   J.~W.~Barrett, R.~J.~Dowdall, W.~J.~Fairbairn, H.~Gomes, F.~Hellmann and R.~Pereira,
  Gen.\ Rel.\ Grav.\  {\bf 43} (2011) 2421,
  arXiv:1003.1886 [gr-qc].
  
  \bibitem{Perez:2012wv}
  A.~Perez,
  Living Rev.\ Rel.\  {\bf 16} (2013) 3,
  arXiv:1205.2019 [gr-qc].

\bibitem{Sahlmann:2001nv}
  H.~Sahlmann, T.~Thiemann and O.~Winkler,
  Nucl.\ Phys.\ B {\bf 606} (2001) 401,
  [gr-qc/0102038].
 
  \bibitem{Sen:2012dw}
  A.~Sen,
  arXiv:1205.0971 [hep-th].

  
\end{thebibliography}
\end{document}